\documentclass[conference]{IEEEtran}
\IEEEoverridecommandlockouts
\usepackage{cite}
\usepackage{amsmath,amssymb,amsfonts}
\usepackage{algorithmic}
\usepackage{graphicx}
\usepackage{textcomp}
\usepackage{xcolor}
\def\BibTeX{{\rm B\kern-.05em{\sc i\kern-.025em b}\kern-.08em
    T\kern-.1667em\lower.7ex\hbox{E}\kern-.125emX}}

\usepackage[pdftex,
            pdfusetitle,
            pdfauthor={Axel Huebl (orcid.org/0000-0003-1943-7141) et al.},
            pdfsubject={A proceedings paper to an invited talk at the Advanced Accelerator Concepts (AAC) Workshop in its 2022 edition.},
            pdfkeywords={simulation, exascale, particle accelerator, particle-mesh, particle-in-cell, laser-plasma, modeling, HPC},
            hidelinks,  % remove colored boxes
            pdfproducer={LaTeX with hyperref},
            pdfcreator={pdflatex}]{hyperref}

\begin{document}

% shorten author lists
\bstctlcite{aac22:BSTcontrol}

% AAC2022
%   https://www.aac2022.org/proceedings/
%   Page Limit: The proceedings should be limited to five pages
%   talk: https://docs.google.com/presentation/d/1fG_uRHXinXkRUhYjVknecvKF4puXvUmfbew6gQ-NfBU/edit

\title{From Compact Plasma Particle Sources to Advanced Accelerators with Modeling at Exascale\\
\thanks{This research was supported by the Exascale Computing Project (17-SC-20-SC), a joint project of the U.S. Department of Energy's Office of Science and National Nuclear Security Administration, responsible for delivering a capable exascale ecosystem, including software, applications, and hardware technology, to support the nation's exascale computing imperative.
This material is based upon work supported by the CAMPA collaboration, a project of the U.S. Department of Energy, Office of Science, Office of Advanced Scientific Computing Research and Office of High Energy Physics, Scientific Discovery through Advanced Computing (SciDAC) program.
This work was supported by the Laboratory Directed Research and Development Program of Lawrence Berkeley National Laboratory under U.S. Department of Energy Contract No. DE-AC02-05CH11231 and by LLNL under Contract DE-AC52-07NA27344.
This research used resources of the Oak Ridge Leadership Computing Facility, which is a DOE Office of Science User Facility supported under Contract DE-AC05-00OR22725, the National Energy Research Scientific Computing Center (NERSC), a U.S. Department of Energy Office of Science User Facility located at Lawrence Berkeley National Laboratory, operated under Contract No. DE-AC02-05CH11231, and the supercomputer Fugaku provided by RIKEN.}
}

% Co-authors based on Title/Scope (accelerator related contribs or research)
\author{\IEEEauthorblockN{%
Axel Huebl\href{https://orcid.org/0000-0003-1943-7141}{\includegraphics[scale=0.19]{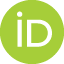}},
R{\'e}mi Lehe\href{https://orcid.org/0000-0002-3656-9659}{\includegraphics[scale=0.19]{orcid.png}},
Edoardo Zoni\href{https://orcid.org/0000-0001-5662-4646}{\includegraphics[scale=0.19]{orcid.png}},
Olga Shapoval\href{https://orcid.org/0000-0003-4003-4507}{\includegraphics[scale=0.19]{orcid.png}},\\
Ryan T. Sandberg\href{https://orcid.org/0000-0001-7680-8733}{\includegraphics[scale=0.19]{orcid.png}},
Marco Garten\href{https://orcid.org/0000-0001-6994-2475}{\includegraphics[scale=0.19]{orcid.png}},
Arianna Formenti\href{https://orcid.org/0000-0002-7887-9313}{\includegraphics[scale=0.19]{orcid.png}},\\
Revathi Jambunathan\href{https://orcid.org/0000-0001-9432-2091}{\includegraphics[scale=0.19]{orcid.png}},
Prabhat Kumar\href{https://orcid.org/0000-0002-8454-7497}{\includegraphics[scale=0.19]{orcid.png}},
Kevin Gott\href{https://orcid.org/0000-0003-3244-5525}{\includegraphics[scale=0.19]{orcid.png}},\\
Andrew Myers\href{https://orcid.org/0000-0001-8427-8330}{\includegraphics[scale=0.19]{orcid.png}},
Weiqun Zhang\href{https://orcid.org/0000-0001-8092-1974}{\includegraphics[scale=0.19]{orcid.png}},
Ann Almgren\href{https://orcid.org/0000-0003-2103-312X}{\includegraphics[scale=0.19]{orcid.png}},\\
Chad E. Mitchell\href{https://orcid.org/0000-0002-1986-9852}{\includegraphics[scale=0.19]{orcid.png}},
Ji Qiang,
Jean-Luc Vay\href{https://orcid.org/0000-0002-0040-799X}{\includegraphics[scale=0.19]{orcid.png}}%
}
\IEEEauthorblockA{
\textit{Lawrence Berkeley National Laboratory}\\
Berkeley (CA), USA\\
\{axelhuebl,jlvay\}@lbl.gov}
\and
\IEEEauthorblockN{%
Alexander Sinn\href{https://orcid.org/0000-0002-4485-971X}{\includegraphics[scale=0.19]{orcid.png}},
Severin Diederichs\href{https://orcid.org/0000-0001-9079-0461}{\includegraphics[scale=0.19]{orcid.png}},\\
Maxence Th{\'e}venet\href{https://orcid.org/0000-0001-7216-2277}{\includegraphics[scale=0.19]{orcid.png}}%
}
\IEEEauthorblockA{\textit{Deutsches Elektronen-Synchrotron (DESY)}\\
Hamburg, Germany}
\and
\IEEEauthorblockN{%
David Grote\href{https://orcid.org/0000-0002-4057-8582}{\includegraphics[scale=0.19]{orcid.png}}%
}
\IEEEauthorblockA{\textit{Lawrence Livermore National Laboratory}\\
Livermore (CA), USA}
\and
\IEEEauthorblockN{%
Luca Fedeli\href{https://orcid.org/0000-0002-7215-4178}{\includegraphics[scale=0.19]{orcid.png}},
Thomas Clark\href{https://orcid.org/0000-0002-0141-8674}{\includegraphics[scale=0.19]{orcid.png}},
Neil Za{\"i}m\href{https://orcid.org/0000-0003-0313-4496}{\includegraphics[scale=0.19]{orcid.png}},
Henri Vincenti\href{https://orcid.org/0000-0002-9839-2692}{\includegraphics[scale=0.19]{orcid.png}}%
}
\IEEEauthorblockA{\textit{LIDYL, CEA-Universit{\'e} Paris-Saclay, CEA Saclay}\\
Gif-sur-Yvette, France}
}

\maketitle

\begin{abstract}
Developing complex, reliable advanced accelerators requires a coordinated, extensible, and comprehensive approach in modeling, from source to the end of beam lifetime.
We present highlights in Exascale Computing to scale accelerator modeling software to the requirements set for contemporary science drivers. %in the community, both in the US and internationally.
In particular, we present the first laser-plasma modeling on an exaflop supercomputer using the US DOE Exascale Computing Project WarpX.
%This includes laser-plasma modeling on an exaflop supercomputer using the US DOE Exascale Computing Project WarpX [1-4] as well as progress of PIConGPU in the OLCF Center for Accelerated Application Readiness (CAAR) project for the same machine, and further projects.
%
Leveraging developments for Exascale, the new DOE SCIDAC-5 Consortium for Advanced Modeling of Particle Accelerators (CAMPA) will advance numerical algorithms and accelerate community modeling codes in a cohesive manner: from beam source, over energy boost, transport, injection, storage, to application or interaction.
Such start-to-end modeling will enable the exploration of hybrid accelerators, with conventional and advanced elements, as the next step for advanced accelerator modeling.
Following open community standards, we seed an open ecosystem of codes that can be readily combined with each other and machine learning frameworks.
These will cover ultrafast to ultraprecise modeling for future hybrid accelerator design, even enabling virtual test stands and twins of accelerators that can be used in operations.
\end{abstract}


\section{Introduction}
Research of plasma-based accelerators has achieved significant milestones over the last decade.
Highlights include achieving nearly 8\,GeV electrons in a single-stage source~\cite{Gonsalves2019}, nC-class electron beams~\cite{Couperus2017}, demonstrating plasma-based FELs~\cite{Wang2021,Galletti2022,Labat2023}, reaching stable proton acceleration of ultra-short, nC-class pulses \cite{Hilz2018} and enabling studies into ultrahigh dose rate radiotherapy~\cite{Kroll2022,Bin2022,Geulig2022}.
As the exploratory aspect of the field benefits significantly from the elucidation of fundamental processes through simulations, transitioning from intriguing sources to scalable accelerators requires universally integrated, quantitatively predictive capabilities for design and operations.

\section{Next-Generation, Advanced Particle Accelerator Modeling}
\begin{figure}[htbp]
  \centerline{\includegraphics[width=\columnwidth]{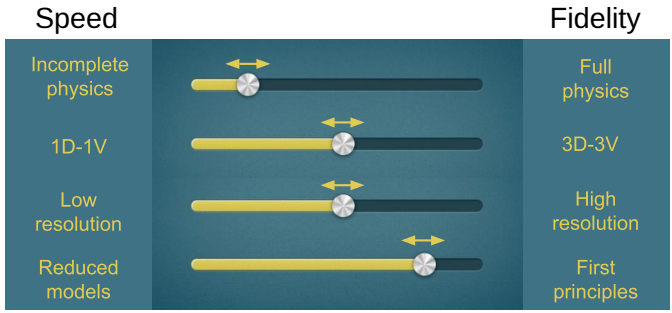}}
  \caption{Preference of speed (time-to-solution) or fidelity (accuracy) of results is determined by implemented algorithms and models in codes.
  Faster codes are preferred for design studies and optimization, more accurate codes for model building, validation and exploration.}
  \label{fig:speedfidelity}
\end{figure}

Computational modeling is an established method in particle accelerator research.
Modeling fulfills tasks from exploratory research for new particle accelerator concepts (including beam physics, model building, validation) to design and optimization of accelerator components.
For these tasks, computational model choices need to strike a balance between running with high fidelity (e.g., for detailed physics studies) or high speed (e.g., for ensemble runs).
Figure~\ref{fig:speedfidelity} provides a schematic overview about possible model choices in a simulations workflow.

In advanced particle accelerator modeling, both extremes of modeling choices often benefit from using leadership-scale supercomputers.
High fidelity modeling is a typical capability computing task in high-performance computing (HPC), requiring large portions of leadership-scale supercomputers at %
the same time to model a single simulation.
On the other hand, fast modeling of ensembles is a typical capacity computing task in high-throughput computing (HTC).

As a consequence of these needs, the community developed more than one ``type'' of particle accelerator modeling code.
Typical choices include fully-kinetic, electromagnetic particle-in-cell codes as the most accurate (and computationally expensive), electrostatic approximations, as well as (partially) fluid-based models.
Most implemented methods these days are explicit algorithms, which iterate forward along either the independent variable of time $t$ or the reference trajectory $s$ of a beam.
Implicit methods are more common in general plasma physics and their research is beneficial for applications with high accuracy requirements, e.g., for long-time stable modeling with high demands on energy conservation.

Scientific drivers for accelerator and beam physics research were recently described as ``Grand Challenges'' in the 2021 Snowmass HEP community exercise~\cite{NagaitsevChallenges2021}:
order-of-magnitude increases in beam intensity, beam quality and phase space density, beam control and beam prediction directly translate into significant needs for computational resolution (grids and no. of particles modeled) and predictive quality (simulate all the particles, conductors, dark currents, many turns, etc.).
In advanced accelerator modeling, specific drivers that require Exascale-supercomputing resources include staged, wakefield-based acceleration for future particle colliders, compact (X)FEL sources, high-field physics experiments (QED) and novel, ultra-intense proton/ion sources.
Closely related scientific domains that benefit from the same modeling capabilities are in high-energy density laser-plasma physics, fusion-energy science, extreme-field science, astrophysical plasmas, light-source modeling, and applications of compact acceleration sources to medical and industrial applications.

\section{Advanced Particle Accelerator Modeling at Exascale}

%\subsection{A Cambrian Explosion of compute architectures}
% maybe

\subsection{Exascale Computing in the US}

The US DOE Exascale Computing Project (ECP) has set its goal to prepare scientists and computing facilities for supercomputers capable of $10^{18}$ double-precision floating point operations per second (1\,ExaFlop/s).
As part of ECP, developed scientific application software aims to reach a domain-specific figure-of-merit that is $50\times$ higher than at project start 7 years earlier.

Targeting the science case of staged wakefield acceleration, \verb|WarpX|\footnote{\href{https://ecp-warpx.github.io}{Homepage: ecp-warpx.github.io}}~\cite{Vay2018,Vay2021,FedeliHuebl2022} is developed in ECP as the successor to the successful \verb|Warp| code~\cite{Vay2013}.
As a particle-in-cell code, the figure-of-merit of \verb|WarpX| in ECP is its parallel particle-and-cell update rate per second.
Recently, \verb|WarpX| completed the ECP goal and even reached a $500\times$ improvement to pre-ECP, due to significant improvements in both hardware and software~\cite{FedeliHuebl2022}.

\subsection{First Runs on Exascale Machines Achieved}

In 2022, the first reported Exascale machine was revealed at Oak Ridge National Lab, called Frontier.\footnote{\url{https://www.olcf.ornl.gov/frontier/}}
Frontier is deployed as part of ECP and is powered by 9,408 compute nodes, each predominantly computing on four AMD MI250X GPUs per node, each with 2 Graphics Compute Dies.
In July 2022, \verb|WarpX|~\cite{FedeliHuebl2022} ran for the first time on the full scale of this Exascale machine.%
\footnote{Also reported at AAC22: Few days later, the PIConGPU~\cite{Huebl2019} team also measured full-scale OLCF Frontier performance results.}

\begin{figure}[htb]
\centering
%\vspace{-3mm}
\includegraphics[width=\columnwidth]{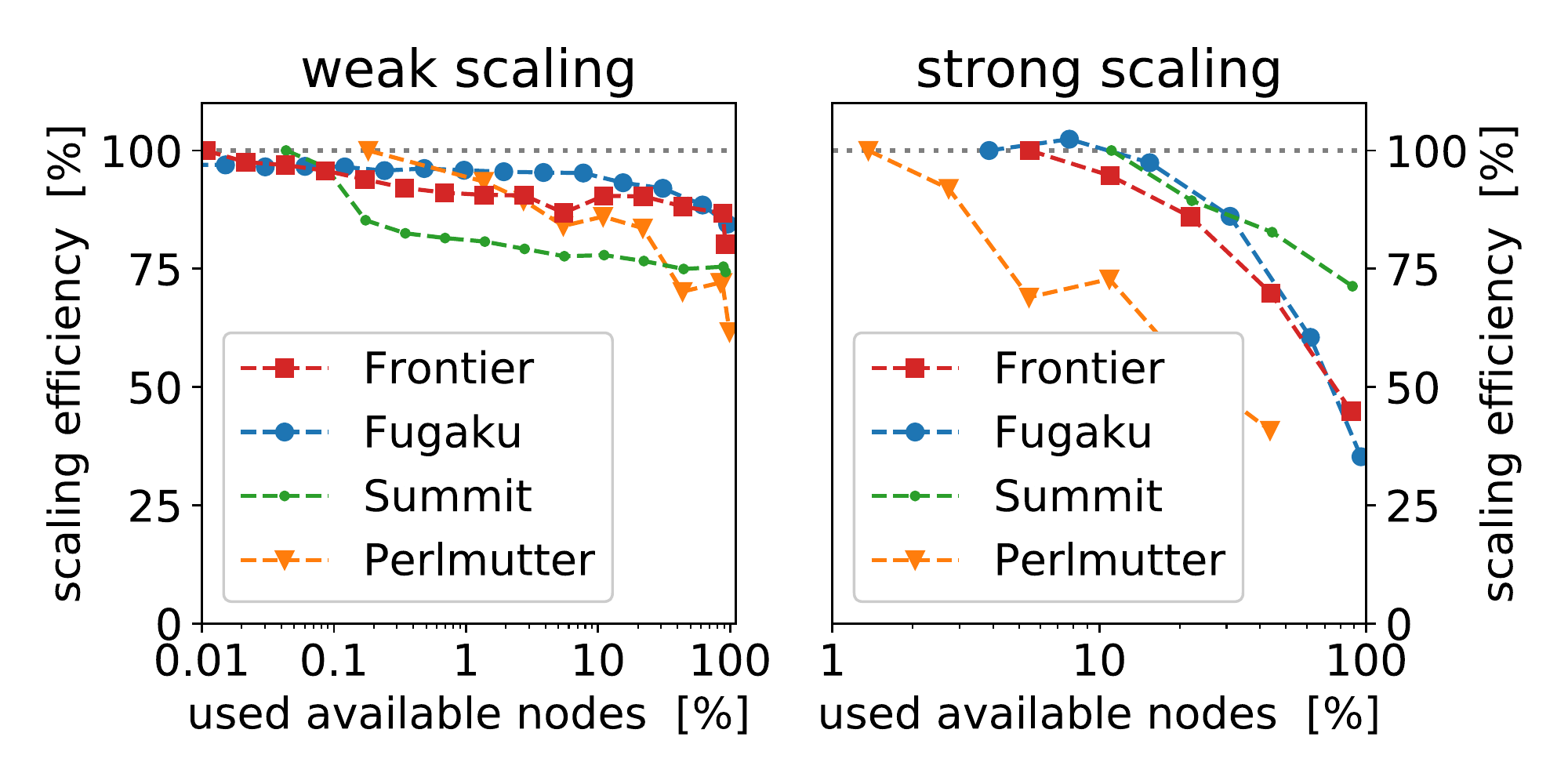}
 \caption{
  Weak and strong scaling of WarpX relative to system size.
  Ideal scaling is the grey dotted line at 100\,\% in both graphs.
  Frontier and Perlmutter measurements were taken prior to their system acceptance dates.
  The network hardware of Perlmutter has since been updated from HPE Slingshot 10 to 11.
  Systems: Frontier (OLCF), Fugaku (Riken), Summit (OLCF), Perlmutter (NERSC).
  Published in~\cite{FedeliHuebl2022}.}
  \label{fig:scaling}
%\vspace{-4mm}
\end{figure}

Over the project years of ECP, \verb|WarpX| has been developed with a performance-portable, single-source GPU-CPU programming model~\cite{Myers2021}.
As a consequence, \verb|WarpX| can run on traditional CPU machines and machines that use hardware accelerators, which are as of today GPUs by three different vendors (Nvidia, AMD, Intel).

\begin{figure*}[htb]
\centering
%\vspace{-3mm}
\includegraphics[width=\textwidth]{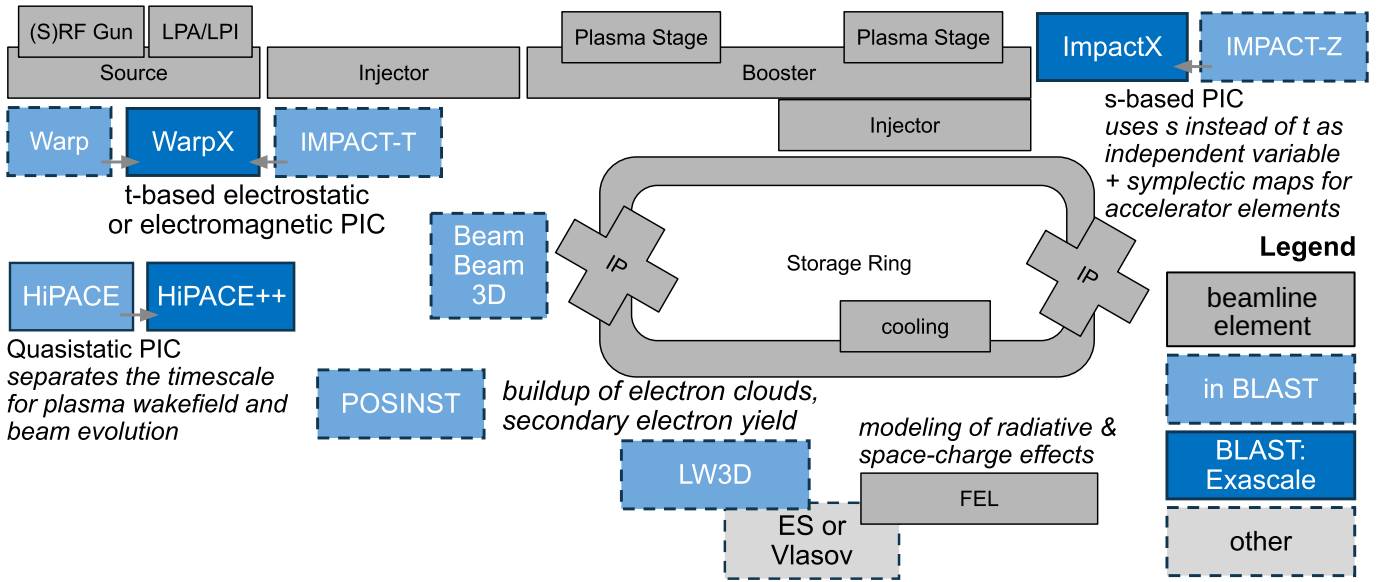}
 \caption{
  Overview of a toy accelerator complex with beamline elements.
  As more BLAST codes are modernized for GPU-support, mesh-refinement and Exascale, projects evolve into the new Exascale code base~\cite{Huebl2022}.
  Abbreviations: Interaction Point (IP), electrostatic (ES), Free Electron Laser (FEL).}
  \label{fig:exablast}
%\vspace{-4mm}
\end{figure*}

Figure~\ref{fig:scaling} from reference~\cite{FedeliHuebl2022} shows the scalability of \verb|WarpX| on some of the largest supercomputers available today.
Highly relevant for the aforementioned challenges in accelerator modeling is the left plot, weak scaling, which shows the code performance when increasing both computational domain and provided parallel computing power on a supercomputer.
\verb|WarpX| achieves close to the ideal scaling over 4-5 orders of magnitude of system size increase.
The plot to the right, strong scaling, shows the decrease in time-to-solution that can be gained by keeping the computational domain constant, but increasing provided parallel computing power.
Due to communication needs in parallel machines, this plot cannot be arbitrarily scaled for complex applications.
Nonetheless, \verb|WarpX| still achieves a remarkable $>50$\,\% efficiency when scaled more than an order of magnitude, before being limited by data communication and GPU/CPU underutilization.

\verb|WarpX| is developed fully in the open, following open science principles~\cite{OpenScienceUNSECO,OpenScienceEU,OpenSource,FreeSoftware} and modern software engineering practices~\cite{LOI_BestPractices}.
Besides running on supercomputers and cloud providers, \verb|WarpX| can also be used on edge and personal computers with Linux, macOS and Windows operating system.
The latter is particularly useful in designing simulation runs in lower dimension and resolution, as well as for development.

\section{Modernizing BLAST for Exascale}

The \verb|WarpX| code is, as was its predecessor \verb|Warp|, part of the BLAST suite of codes.\footnote{\url{https://blast.lbl.gov}}
BLAST, originally standing for ``Berkeley Lab Accelerator Simulation Toolkit'', was renamed in 2021 to ``\textbf{B}eam, P\textbf{l}asma \& \textbf{A}ccelerator \textbf{S}imulation \textbf{T}oolkit''.
This reflects the grown community that spans maintainers, contributors and collaborators from many international institutions, and applies the BLAST codes to a growing number of applications.
For instance, a code primarily developed outside of LBNL is the quasi-static \verb|HiPACE++| code~\cite{Diederichs2022}, which is maintained by DESY.
Another code is \verb|GEMPIX|, developed at IPP Garching for magnetic confinement fusion plasmas.

Leveraging the success and routines of \verb|WarpX|, other existing BLAST codes are being transitioned to Exascale~\cite{Huebl2022} as well.
Figure~\ref{fig:exablast} shows an overview of the new GPU-capable, mesh-refinement (MR)-enabled codes for accelerator modeling and prominent application.
During this transition, existing codes are rewritten from Fortran to C++ for core compute routines and modern data structures.
A common input layer is developed, which is standardized in the Python API \verb|PICMI| (particle-in-cell modeling interface).\footnote{\url{https://picmi-standard.github.io}}
Time-based ($t$) implementations of the codes \verb|Warp| and \verb|IMPACT-T|~\cite{impactt} are combined in the new electromagnetic and electrostatic \verb|WarpX| code.
Beam-dynamics codes along a reference trajectory $s$ as the independent variable, such as \verb|IMPACT-Z|~\cite{impactz} and $s$-based \verb|Warp| modules, are modernized in the new code \verb|ImpactX|~\cite{Huebl2022} and continue to include collective effects.
Common particle-in-cell routines are shared via the Accelerated BLAST Recipes (\verb|ABLASTR|) library, generalizing \verb|WarpX| routines~\cite{Huebl2022}.

The Open Standard for Particle-Mesh Data (\verb|openPMD|) is used in BLAST codes for data compatibility~\cite{openPMDstandard}.
A high-performance, \verb|openPMD| C++ and Python reference implementation, co-developed by LBNL and HZDR/CASUS, is used in massively parallel codes and in data analysis~\cite{openPMDapi}.
Recently, data streaming techniques were developed to transition traditional post-processing scripts to online, massively parallel data analysis workflows, which can be co-located with simulations and enable rapidly prototyping for scientific \textit{in-transit} analysis~\cite{Poeschel2022}.

\section{Seeding a Community Ecosystem}

From experience over the last years developing community standards such as \verb|openPMD|, \verb|PICMI|, the BLAST toolkit, or the \verb|PIConGPU| software stack~\cite{Huebl2019} - open standardization and modular, open source software collaborations emerge as the necessarily efficient way forward.
From such seed projects, the US DOE SCIDAC-5 Collaboration for Advanced Modeling of Particle Accelerators (CAMPA)~\cite{CAMPA} will support advancement of numerical algorithms and accelerate community modeling codes in a cohesive manner: from beam source, over energy boost, transport, injection, storage, to application or interaction.

Making the existing particle accelerator modeling ecosystem compatible and interdependent will be beneficial towards the goal of predictive start-to-end modeling~\cite{LOI_ecosystem}.
As a community, we should expect and ready models and codes for the exploration of ``hybrid'' accelerators, with conventional and advanced elements, as the next step for advanced accelerator modeling.
Following open community standards, one can initiate an open ecosystem of codes that can be readily combined with each other and machine learning frameworks~\cite{Huebl2022}, towards enabling virtual test stands and twins of accelerators that can be used in operations.

\section*{Acknowledgment}

%``R. B. G. thanks$\ldots$''.
%Put sponsor acknowledgments in the unnumbered footnote on the first page.
This research used the open-source particle-in-cell code \verb|WarpX| \url{https://github.com/ECP-WarpX/WarpX}, primarily funded by the US DOE Exascale Computing Project.
Primary \verb|WarpX| contributors are with LBNL, LLNL, CEA-LIDYL, SLAC, DESY, CERN, and TAE.
We acknowledge all \verb|WarpX|, \verb|HiPACE++|, \verb|ImpactX| and \verb|openPMD| contributors.
Slides of the plenary presentation are available under~\cite{slides}.

% References
%\bibliographystyle{IEEEtran}
%\bibliography{biblio}


\end{document}